\documentclass[review,number,sort&compress]{elsarticle}

\usepackage{lineno, hyperref}
\usepackage{amsmath}
\usepackage{amssymb}
\usepackage{color}

\begin{document}
\begin{frontmatter}

\title{AXEL: High-pressure Xe gas TPC for BG-free $0\nu2\beta$ decay search}


\author[hepKyoto]{S.~Obara\corref{mycorrespondingAuthor}}
\cortext[mycorrespondingauthor]{Corresponding author}
\ead{obara.shuhei.68c@st.kyoto-u.ac.jp}

\author[hepKyoto]{S.~Ban} 
\author[hepKyoto]{M.~Hirose} 
\author[hepKyoto]{A.~K.~Ichikawa} 
\author[hepKyoto]{T.~Kikawa} 
\author[hepKyoto]{K.~Z.~Nakamura} 
\author[hepKyoto]{T.~Nakaya} 
\author[hepKyoto]{S.~Tanaka} 
\author[hepKyoto]{M.~Yoshida}

\address[hepKyoto]{Department of Physics, Kyoto University, Kyoto 606-8502, Japan}

\author[kyticr]{Y.~Iwashita} 
\address[kyticr]{Advanced Research Center for Beam Science, Kyoto University, Kyoto 611-0011, Japan}

\author[icrr,ipmu]{H.~Sekiya} 
\address[icrr]{Kamioka Observatory, Institute for Cosmic Ray Research, the University of Tokyo, Hida 506-1205, Japan}

\author[icrr,ipmu]{Y.~Nakajima} 
\address[ipmu]{Kavli Institute for the Physics and Mathematics of the Universe (WPI), the University of Tokyo, Kashiwa 277-8568, Japan}

\author[rcns]{K.~Ueshima} 
\address[rcns]{Research Center for Neutrino Science, Tohoku University, Sendai 980-8578, Japan}

\author[kobe]{K.~Miuchi} 
\author[kobe]{K.~D.~Nakamura} 
\address[kobe]{Department of Physics, Kobe University, Kobe 657-8501, Japan}

\author[yokohama]{A.~Minamino} 
\address[yokohama]{Yokohama National University, Faculty of Engineering, Yokohama 240-8501, Japan}

\author[kek]{T.~Nakadaira}
\author[kek]{K.~Sakashita}
\address[kek]{Institute of Particle and Nuclear Studies, High Energy Accelerator Research Organization (KEK), Tsukuba 305-0801, Japan}

\begin{abstract}
  AXEL is a high-pressure xenon gas time projection chamber for neutrinoless double-beta decay ($0\nu2\beta$) search.
  The AXEL has a unique readout system called ELCC which has a cellular structure and photosensors to detect electroluminescence light produced by ionization electrons.
  We demonstrated the performance of the ELCC with a small prototype detector ({AXEL-HP10L}).
  The obtained energy resolution corresponds to $0.82-1.74\%$ (FWHM) at the $0\nu2\beta$ Q-value.
  We are constructing a new prototype ({AXEL-HP180L}) in order to study the energy resolution at the Q-value of $0\nu2\beta$ 
  with a new design of ELCC with unit structure,
  newly developed electronics board, field-shaping electrodes, and 
  Cockcroft-Walton-type high voltage power supply.
  For a future 1-ton scale large AXEL detector, we are developing new background-reduction techniques; topology identification with machine learning, positive-ion detection, and active-shield options.
  
\end{abstract}
 
\begin{keyword}
  Double Beta Decay; Majorana Neutrino; Gas Detector; Xenon;
\end{keyword}
\end{frontmatter}


\section{Introduction}
The neutrinoless double-beta decay ($0\nu2\beta$) has an important role to reveal the neutrino mysteries; mass hierarchy, absolute mass, and Majorana nature.
Double-beta decay ($2\nu2\beta$) with two-neutrinos and two-electrons emission is allowed for certain nuclei in the standard model.
If neutrino is Majorana type ($\nu = \nu^C$), $0\nu2\beta$ is also allowed through the Majorana-mass term.
The half-life of $0\nu2\beta$ is calculated as $(T_{1/2}^{0\nu})^{-1} = G^{0\nu} |M^{0\nu}|^2 \langle m_{\beta\beta}\rangle^2$, where $G^{0\nu}$ is a phase-space factor, $M^{0\nu}$ is a nuclear matrix element, and $\langle m_{\beta\beta}\rangle^2 = |\Sigma_i U^2_{ie} \nu_i|^2$ is an effective neutrino mass.
The stringest limit is given by the KamLAND-Zen experiment as ${T_{1/2}^{0\nu}>1.07\times10^{26}}$~yr corresponding to ${\langle m_{\beta\beta} \rangle < 61 - 165}$~meV~\cite{PhysRevLett.117.082503}, which is close to the expected value in case of the inverted-mass hierarchy.
In order to explore the whole inverted-mass hierarchy region, the future $0\nu2\beta$ experiments are aiming to reach $T_{1/2}^{0\nu} \gtrsim 1.0\times10^{28}$~yr.
The keys of high-sensitive $0\nu2\beta$ search are (i) a large amount of double-beta nuclei, (ii) good energy resolution, and (iii) background-rejection power.

A high-pressure xenon-gas time projection chamber (TPC) can have three advantages mentioned above with ${}^{136}{\rm Xe}$ as the double-beta nuclei.
The natural abundance is 8.9\% and it can be enriched to $\sim$ 90\%.
The Q-value is 2,458~keV.
There are three on-going projects; NEXT experiment~\cite{Alvarez:2011my}, PandaX-III experiment~\cite{Chen:2016qcd}, and AXEL experiment.
In this paper, we report the current status of the AXEL project.

\section{Overview of the AXEL Detector}
The AXEL ({\bf A} {\bf X}enon {\bf E}lectro{\bf L}uminescence) detector contains a 6 to 8 bar pure xenon gas and utilities the electroluminescence (EL) process to detect ionization electrons and to measure the energy and tracks of two-electrons from $0\nu2\beta$~\cite{BAN2017185}.

\begin{figure}[htb]
  \centering
  \includegraphics[width=1.0\linewidth]{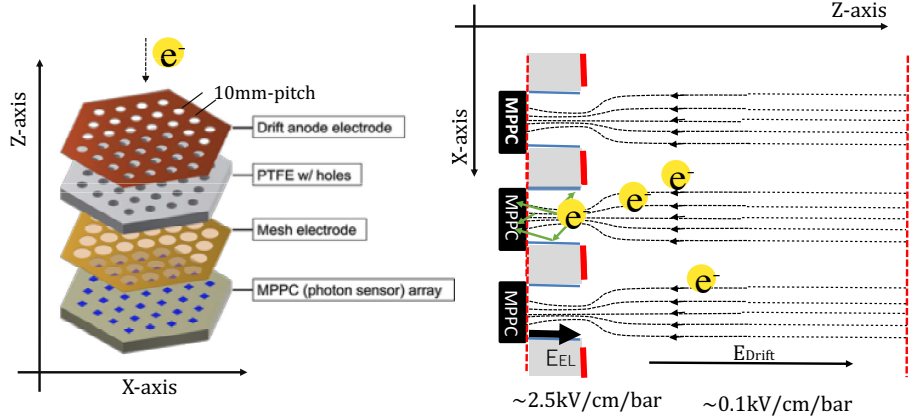}
  \caption{Schematic view of the ELCC tracking plane of the AXEL detector. ELCC consists of four layers; anode copper plate, PTFE insulator, GND mesh electrode, and MPPC array. 
  The copper plate and PTFE plate have 5.5-mm diameter holes in 10~mm pitch and MPPC's are placed to the corresponding positions.
  The ionization electrons are attracted into the holes and produce EL-photons, which are then detected by the MPPC.}  
  \label{fig_elcc}
\end{figure}

In a TPC, ionization electrons produced along the $\beta\beta$-ray tracks are drifted towards the anode by an electric field ($E_{\rm drift}\sim$ 0.1 kV/cm/bar) and then detected by a tracking plane there.
The AXEL detector has a unique tracking plane called ``ELCC'' ({\bf E}lectroluminescence {\bf L}ight {\bf C}ollection {\bf C}ell) shown in Figure~\ref{fig_elcc}.
The ELCC consists of four layers.
The top copper plate is an anode to produce the drift field.
The second layer is a rigid PTFE insulator which supports the electrodes.
These two layers have holes of 5.5~mm in diameter placed with 10-mm pitch hexagonal pattern.
The third layer is a mesh electrode which is set to be GND.
A voltage is applied between the top copper plate and the GND mesh to form the electroluminescence electric-field ($E_{\rm EL} \sim 2.5$~kV/cm/bar) collecting the ionization electrons into each hole.
Such ionization electrons are accelerated in a hole and produce the electroluminescence photons (EL-photons).
Finally, the bottom layer is an array of the multi-pixel photon counters (MPPC) to detect the EL-photons.

The photon generation by electroluminescence is a proportional process. 
Thanks to the small W-value of xenon ionization and the linear-amplification, the expected energy resolution is as good as 0.5\% (FWHM) at the Q-value of $0\nu2\beta$ (2,458~keV), which is good enough to distinguish the $0\nu2\beta$ signal from the $2\nu2\beta$ background.

\section{Performance Demonstration by Prototype Detectors}
\begin{figure}[htb]
  \centering
  \includegraphics[width=1.0\linewidth]{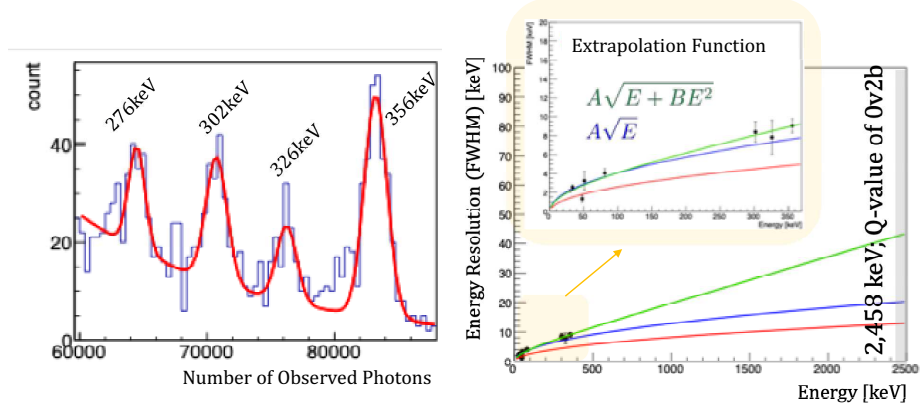}
  \caption{Energy spectrum of ${}^{133}{\rm Ba}$ source run (Left) and the energy resolution (Right) of {AXEL-HP10L}. In the right figure, green and blue lines are the extrapolation function of $A\sqrt{E+BE^2}$ and $A\sqrt{E}$, respectively. Estimated energy resolution is $0.82 - 1.74$\% (FWHM) at the Q-value. Details of these studies are described in the references~\cite{BAN2017185,yoshida_masashi_2018_1300681}. (Color online.)
  }
  \label{fig_resolution}
\end{figure}

A 10-L size prototype detector ({AXEL-HP10L}) was built to demonstrate and optimize ELCC.
The obtained energy resolution is 2.54\% (FWHM) at 356~keV and corresponds to $0.82 - 1.74\%$ (FWHM) at the Q-value of $0\nu2\beta$ with extrapolation as shown in Figure~\ref{fig_resolution}~\cite{BAN2017185,yoshida_masashi_2018_1300681}.

As a next step, we are constructing a 180-L prototype detector ({AXEL-HP180L}) to check the energy resolution at the Q-value; the {AXEL-HP10L} is too small to observe electrons having energy around the Q-value.

\begin{figure}[htb]
  \centering
  \includegraphics[width=1.0\linewidth]{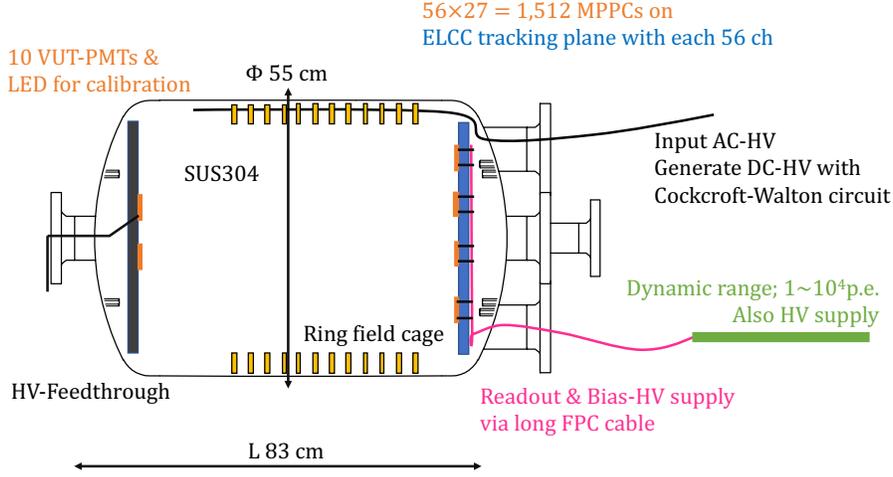}
  \caption{Schematic side view of the {AXEL-HP180L} prototype.
  }
  \label{fig_hp180l}
\end{figure}

Figure~\ref{fig_hp180l} shows the schematic view of the {AXEL-HP180L}.
The ELCC plane is divided into 27-units, each of which has a 7~cm $\times$ 8~cm parallelogram shape with 56 cells and MPPCs.
The total fiducial area is 438.3~mm in diameter.
The ELCC plane can be easily enlarged for larger detector thanks to the unit structure. 
Each unit is connected to a newly developed electronics board via a long flexible-print-cable.
The board records the MPPC waveforms at 5~MHz sampling with a dynamical range of $1 - 10^4$ photo-electrons (p.e.) per 1~$\mu$sec corresponding to a few keV to about 2.5~MeV.
It also supplies the bias-voltage to each MPPC~\cite{Tanaka_2017}.

At the other side, a cathode mesh, tens of photomultiplier tubes and LED's are mounted.
The LED's are used to measure the response of MPPCs.

\begin{figure}[htb]
  \centering
  \includegraphics[width=1.0\linewidth]{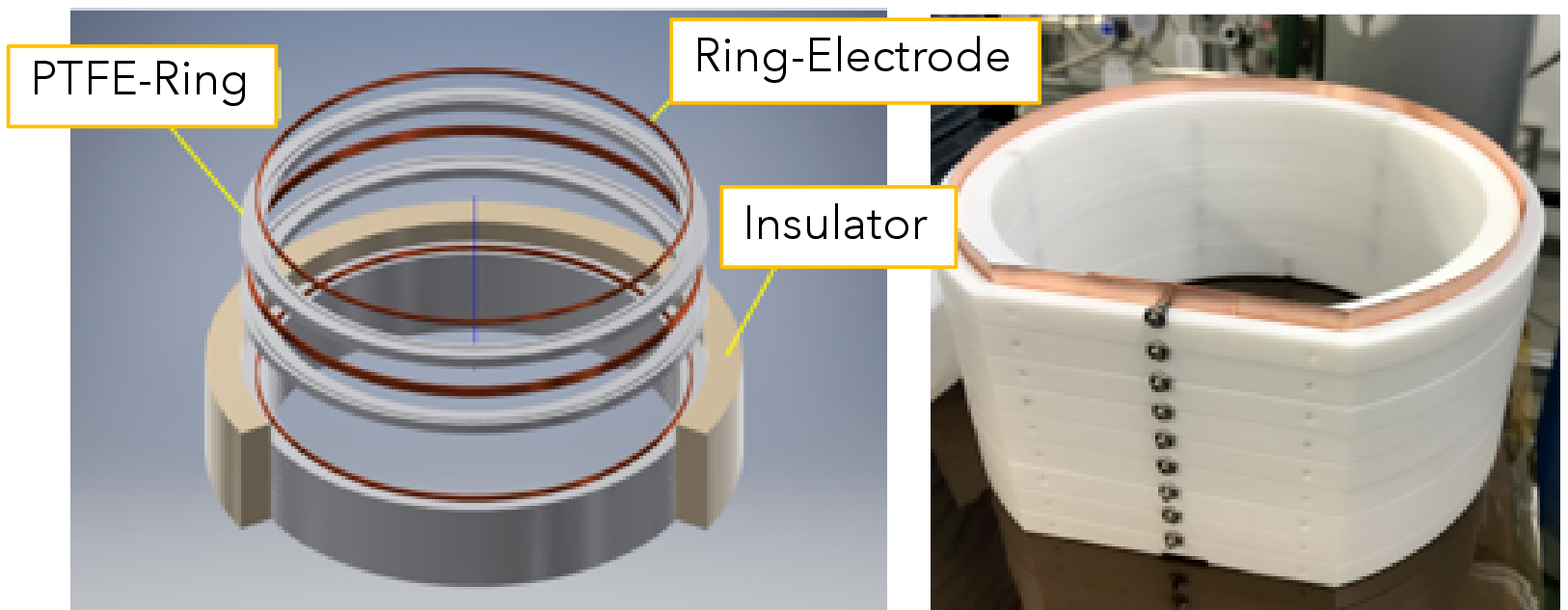}
  \caption{Prototype of a field cage and PTFE covers to prevent discharge between the chamber wall and the electrodes. 
  (Color online.)
  }
  \label{fig_fieldcage}
\end{figure}

The drift electric field is formed by a Cockcroft-Walton circuit that generates the high DC voltage in the chamber and by field-shaping electrodes~\cite{yoshida_masashi_2018_1300681}.
In order to prevent a discharge from the electrodes, a PTFE insulator sandwiches the electrodes and fills the space between the wall of the chamber and electrodes as shown in Figure~\ref{fig_fieldcage}.
This PTFE also plays as a reflective surface for the scintillation light and increase the number of photons detected by the photomultiplier tubes.

The data taking with the {AXEL-HP180L} detector will start soon with three ELCC units ($56\times3 = 168$ MPPC channels).
The full ELCC units will be installed by the end of 2019.

\section{R\&D's of new Technologies}

\subsection*{Topology Identification}
Although the high-pressure xenon-gas TPC has a good energy resolution, photo-electrons produced by 2,447~keV gamma-rays from ${}^{214}{\rm Bi}$ cannot be distinguished from the signal of ${}^{136}{\rm Xe}$-$0\nu2\beta$ by the energy~\cite{Alvarez:2011my}.
However, the gas TPC provides the track pattern information, which can be used and has a possibility to discriminate background.
The gamma-ray photoelectric background has only one track and one blob at the end, on the other hand, the $0\nu2\beta$ signal has two tracks starting from one point and two blobs at ends.
Recent machine-learning algorithms are useful for such track-image identification.
We developed a neural-network algorithm based on DenseNet~\cite{Huang_2017_CVPR}.
Its deep-learning identification is expected to reduce the background rate from ${}^{214}{\rm Bi}$ in the very radio-pure copper vessel to 0.8~cnt/yr for a 1-ton xenon gas TPC experiment. 
The signal efficiency is 27\%, but the background contamination can be suppressed to 0.0004\%.
The sensitivity to the half-life of $0\nu2\beta$ is expected as $T^{0\nu}_{1/2} > 1\times 10^{27}$~yr for six-years operation.
However, in order to reach the normal-hierarchy region, we have to reduce the background contamination further.

\subsection*{Positive Ion Detection}
Positive ions (${\rm Xe}^+$, ${\rm Xe}_2^+$, or ${\rm Xe}_3^+$), which are generated along the $0\nu2\beta$ tracks, slowly drift to the cathode plane.
Since the positive ions have smaller diffusion constant compared to the ionization electrons, the positive-ion tracking information may help the pattern identification.
The basic studies is also conducted in NEXT experiment~\cite{lior}.

To detect the positive ions, we use the secondary-electron emission from the cathode electrode.
If the ionization energy of the positive ion is higher than the work function of electrode, a secondary electron can be emitted even if the kinetic energy of the ion is small.

\begin{figure}[htb]
  \centering
  \includegraphics[width=0.8\linewidth]{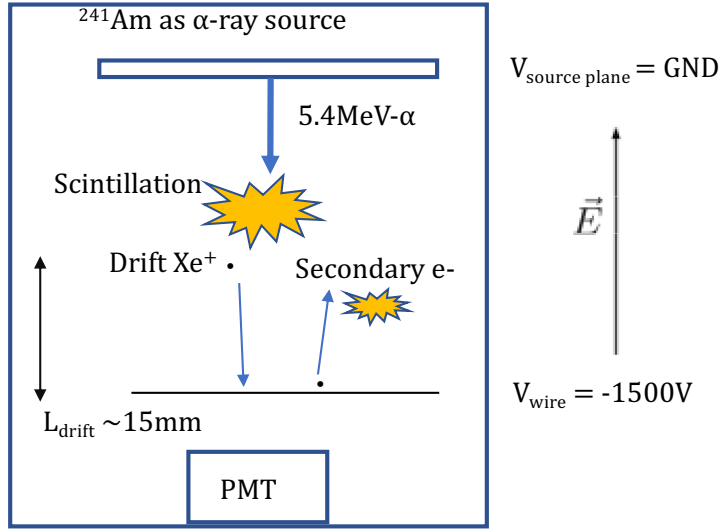}
  \caption{Schematic view of the test setup for positive-ion detection in xenon-gas TPC. 
  A top plate is set to be GND, and the tungsten-wire cathode is -1500~V.
  ${}^{241}{\rm Am}$ alpha source is mounted at the top. 
  The scintillation light and EL-photons by secondary electrons are detected by a photomultiplier tube at the bottom.
  }
  \label{fig_positiveion}
\end{figure}

\begin{figure}[htb]
  \centering
  \includegraphics[width=1.0\linewidth]{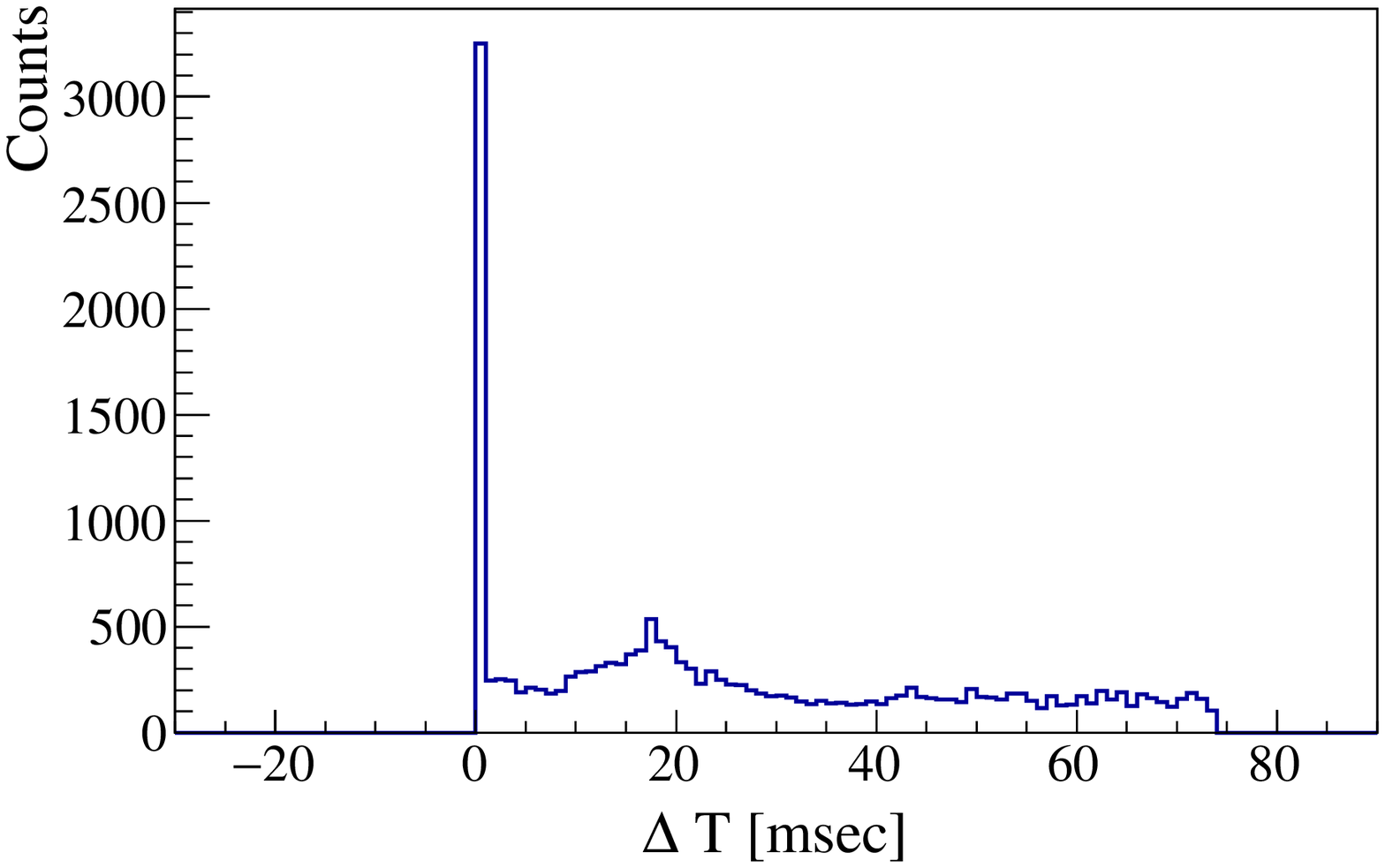}
  \caption{Distribution of pulse timing after the scintillation signal. The peak at $T=0$ corresponds to the scintillation signal by alpha-rays. The plot is obtained by accumulating many alpha-ray events.}
  \label{fig_deltat}
\end{figure}

For a basic study of the positive-ion detection, a $\phi=20$-$\mu$m tungsten wire was set as the secondary-electron source in a chamber filled with 1-atm xenon gas, as shown in Figure~\ref{fig_positiveion}.
The wire is surrounded by a cylindrical gas region of 2-cm diameter and 3.3-cm height.
An electric field is produced by applying $-1500$~V to the tungsten-wire cathode against the top GND plate.
An ${}^{241}{\rm Am}$ alpha-ray source is mounted at the top plate.
An alpha-ray stops with its short range of about 20~mm and makes scintillation, electrons, and positive ions.
After $\sim$ 15~mm long drift, the positive ions reach the wire and may induce secondary electrons from the surface of the wire cathode.
The secondary electrons would then emit EL-photons.
A photomultiplier tube is placed at the bottom of the chamber to detects EL-photons.
The drift time of the positive-ions is estimated to be $10-20$~msec from the ion mobility of 1.5~mm/msec at 200~V/cm/bar~\cite{2007NIMPA.580...66N}.
Figure~\ref{fig_deltat} shows the obtained distribution of pulse timing after the scintillation signal.
The distribution seems to have a slight excess at around 18~msec, which is consistent with our expectation of the drift time of ions.
Further study will be conducted to confirm the ion detection.

\subsection*{Active Shield and Chamber Vessel}
Significant amount of background is expected from ${}^{214}{\rm Bi}$ gamma-ray in the chamber vessel.
For the reduction of such background, we have some optional plans.

One is a thin oxygen-free copper vessel surrounded by pressurised water.
The pressurised water supports the thin chamber filled with high-pressure gas from the outside.
The water can also be used as a water-Cerenkov outer detector for a cosmic-muon veto.
Also, liquid-scintillator or gadolinium-loaded water-Cerenkov detectors are other candidates for the outside medium.

Another new idea is an active chamber vessel made of a polyethylene-naphthalate resin.
A polyethylene naphthalate has a blue photon emission (425~nm of wavelength) and can work as scintillator~\cite{Nakamura_2011}.
If this active vessel is realized, most of the ${}^{214}{\rm Bi}$ backgrounds from the vessel can be identified in an offline analysis by the delayed-coincidence method using the sequential alpha-decay of ${}^{214}{\rm Po}$ as a delayed event~\cite{Obara:2017ndb}.
We succeeded to make a spherical-shape vessel of the polyethylene-naphthalate resin with 10~mm thickness and 300~mm diameter by the injection molding.
However, generally, plastic materials emit a lot of outgas, that is a serious problem for the pure xenon gas TPC.
We will evaluate the outgassing rate and survey various active materials.

\section{Summary}
AXEL is a high-pressure gaseous ${}^{136}{\rm Xe}$ TPC for $0\nu2\beta$ search with a unique tracking-plane ELCC. 
We established the basic design of the ELCC and are constructing a prototype detector called {AXEL-HP180L} to demonstrate a performance at the ${}^{136}{\rm Xe}$ Q-value. 
New background-rejection technologies are also being developed; machine learning, positive-ion detection, and active-shield chamber.
We will start to run the {AXEL-HP180L} by the end of 2019.

\section*{Acknowledgment}
We would like to thank the VCI 2019 organization committee for this opportunity to present the AXEL project.

The AXEL project is supported by the Grant-in-Aid for Japan Society for the Promotion of Science Fellows Grant Number 15H02088, 16J09462, 17K18777, 17J00268, 18J13957, 18J00365, 18J20453, 18K13567, and 18H05540. 

We also thank the Institute for Cosmic Ray Research and the Kamioka underground laboratories, University of Tokyo for their support of our project.

\section*{Reference}

\end{document}